# Diamond quantum sensing at record high pressure up to 240 GPa

Qingtao Hao[1,2,#], Ze-Xu He[3,4,#], Na Zuo[1,2,#], Yang Chen[3,#], Xiangzhuo Xing[1,2,*], Xiaoran Zhang[1,2,*], Xinyu Zhuang[1,2], Zhixiang Shi[5], Xin Chen[1,2], Jian-Gang Guo[3], Gang-Qin Liu[3,4,*], Xiaobing Liu[1,2,6*] & Yanming Ma[6,7]

[1]*Key Laboratory of Quantum Materials under Extreme Conditions in Shandong Province, School of Physics and Physical Engineering, Qufu Normal University, Qufu 273165, China.*

[2]*Laboratory of High Pressure Physics and Material Science (HPPMS), Advanced Research Institute of Multidisciplinary Sciences, Qufu Normal University, Qufu 273165, China.*

[3]*Beijing National Laboratory for Condensed Matter Physics and Institute of Physics, Chinese Academy of Sciences, Beijing, 100190, China.*

[4]*School of Physical Sciences, University of Chinese Academy of Sciences, Beijing, 100190, China.*

[5]*School of Physics, Southeast University, Nanjing 211189, China.*

[6]*State Key Laboratory of Superhard Materials, Key Laboratory of Material Simulation Method and Software of Ministry of Education, College of Physics, Jilin University, Changchun 130012, China*

[7]*International Center of Future Science, Jilin University, Changchun 130012, China*

[#]*These authors contributed equally to this work.*

[*]*Email: xzxing@qfnu.edu.cn, xiaoran_zhang@qfnu.edu.cn, gqliu@iphy.ac.cn, xiaobing.phy@qfnu.edu.cn*

# Abstract

Quantum sensing utilizing nitrogen-vacancy (NV) centers in diamond has emerged as a transformative technology for probing magnetic phase transition[1-4], evidencing Meissner effect of superconductors[1,5-9], and visualizing stress distribution[3,9] under extreme conditions. Recent development in NV configurations and hydrostatic environments have raised the operational pressures of NV centers to 140 GPa[2,6,10,11], but substantial challenges remain in extending sensing capabilities into multi-megabar range, critical for research in hydrogen-rich superconductors like La-Sc-H ($T_c$ of 271-298 K at 195-266 GPa)[12] and evolution of minerals near Earth's core[13]. Here we report the fabrication of shallow NV centers through ion implantation followed by high-pressure and high-temperature (HPHT) annealing, leading to increased density, improved coherence, and mitigated internal stresses, a pre-requisite for reducing their degradation under compression. This NV magnetometry enable breakthrough of pressure capabilities exceeding 240 GPa, constrained by structural integrity of the 50 μm diamond anvils, suggesting that the untapped pressure limit may enable further advancements with smaller cutlets or more robust diamonds. We present compelling evidence of the Meissner effect and trapped flux at record-high pressure of 180 GPa for superconducting transition in elemental titanium (Ti) as benchmark, establishing a solid foundation for high-pressure magnetometry in exploring complex quantum phenomena at previously unreachable pressures.

Extreme high-pressure conditions up to multiple megabars play vital roles for advancing research in fields such as condensed matter physics, material science and geoscience. With the development of high-pressure technologies, particularly diamond anvil cell (DAC), has enabled *in-situ* investigations of structural phase transitions[14,15], as well as optical[16] and electrical transport properties[17]. However, magnetic detection remains one of the principal challenges in the community. Conventional techniques, such as superconducting quantum interference devices (SQUIDs) and a.c. magnetic susceptibility, are enslaved to the weak signals generated by tiny samples (typically several to tens of micrometers in diameter) and the significant spatial separation from the detection apparatus within DAC. These constraints lead to significant scientific challenges, including ongoing debates regarding the origins of Earth's magnetism[13], where core pressures reach approximately 360 GPa, and a lack of compelling evidence supporting the Meissner effect in numerous pressure-induced superconducting transitions of hydrogen-rich compounds (e.g., $H_3S$ at 150 GPa[18], $LaH_{10}$ at 170 GPa[19,20], $CaH_6$ at 172 GPa[21,22], $YH_9$ at 201 GPa[23]), which are considered potential candidates for high-temperature and even room-temperature superconductors (such as La-Sc-H[12], 271-298 K at 195-266 GPa). Therefore, an urgent need for advanced approaches to magnetic detection at multiple megabar pressures exists, as addressing this challenge is vital for understanding fundamental physical phenomena, realizing the potential of next-generation superconductors and novel pressure-stabilized magnetic materials.

Quantum sensing based on nitrogen-vacancy (NV) centers in diamond offers an innovative approach for detecting magnetic fields[24-26], temperature[27,28], and pressure[3,29], featuring exceptional sensitivity and spatial resolution. Each NV center acts as an atomic-scale single-photon defect in diamond lattice[30,31], consisting of a substitutional nitrogen atom adjacent to a vacancy, with an electronic ground state with spin $S = 1$. Under external perturbations, the spin sublevels undergo energy splitting or shifts, which can be read out through optically detected magnetic resonance (ODMR), which facilitates highly sensitive measurements[30,32]. Initial demonstrations[29] of NV magnetic sensing at high pressures of 0-60 GPa in 2014 paved the way for *in-situ* investigations of pressure-induced magnetic phase transitions in magnetic materials[1-4], superconducting transitions[1,5-9], as well as pressure imaging[3,9]. Recent progress has expanded the operational pressure limits of NV centers to the megabar range by optimizing hydrostatic pressure environments, utilizing micron-sized diamond particles[10] and micropillars on diamond anvil culets[11], thereby affirming their potential as reliable magnetic sensors. An important achievement is the development of shallow NV centers on [111]-oriented diamond anvils that modulate uniaxial stress along the NV axis while preserving symmetry under compression, thereby enabling magnetic sensing capabilities up to 140 GPa and facilitating the observation of the magnetic phase transition of $Fe_3O_4$ at 128 GPa[2] and the Meissner effect in $CeH_9$ at 137 GPa[6]. The refinement of NV configurations and their hydrostatic environments is essential for advancing operational pressure limits in magnetometry. Nevertheless, the pursuit of capabilities in the multiple megabar range remains substantial challenges, emphasizing the critical need for novel effective strategies and a comprehensive understanding of the intrinsic properties of NV centers.

Conventional fabrication methods for NV centers exist critical obstacles, including the inevitable lattice damage caused by nitrogen implantation and electron irradiation[33-35], as well as the potential graphitization of diamond during annealing processes exceeding 800 °C[33,36]. Furthermore, the presence of residual substitutional nitrogen atoms (P1 centers) further complicates these issues[24,36]. These factors collectively lead to a reduction in the NV density, particularly affecting shallow NV centers, and generate a chaotic spin bath[24], which significantly undermines the stability of NV centers at pressures and diminishes their detection sensitivity. Addressing these challenges is imperative for enhancing the performance of NV centers in magnetometry. Here we report the successful fabrication of shallow NV centers on diamond anvils through a strategic process of ion implantation followed by HPHT annealing treatment, demonstrating significant enhancements in density, coherence properties, and stress optimization, as well as improved diamond transparency. These advancements enable long-anticipated NV sensors to conduct research at extreme pressures exceeding 240 GPa, while still maintaining exceptional sensitivity, corresponding to the critical failure point of 50 μm diamond anvils. Upon this NV center, we provided compelling evidence of Meissner effect and trapped flux at a record-high pressure of 180 GPa in elemental Ti superconductor[37-40]. Our findings strongly suggest that the full potential of NV sensing remains untapped, indicating that forthcoming advancements utilizing smaller culets or more robust diamonds could lead to unprecedented breakthroughs in high-pressure magnetometry.

We synthesized high quality type-IIa diamond single crystals using HPHT techniques and cut along the (111) crystallographic plane to produce a 3×3×1 $mm^3$ diamond slide for depth measurement of the fabricated shallow NV centers. As illustrated in Fig. 1a, nitrogen ions were implanted into the diamond surface at a fluence of 3×10$^{14}$ $cm^{-2}$, with a subsequent characterization revealing an implantation depth of approximately 50 nm (Fig. 1b) using secondary ion mass spectrometry (SIMS) and three-dimensional (3D) microscopy. HPHT annealing enhanced the aggregation of nitrogen atoms and vacancies to promote NV center formation while effectively suppressing diamond graphitization, enabling an increased annealing temperature of 1500 °C. Figure 1c shows Raman spectra for samples annealed at 800 °C under vacuum and 1500 °C at 6 GPa. The HPHT-annealed diamond exhibits a peak at 1332.1 cm$^{-1}$, close to the ideal diamond peak of 1332 cm$^{-1}$, indicating effective relief of tensile stress from ion implantation. This is further corroborated by continuous-wave ODMR spectra, which show a reduction of approximately 2 MHz in zero-field splitting for HPHT-fabricated NV centers (Extended Data Fig. 1).

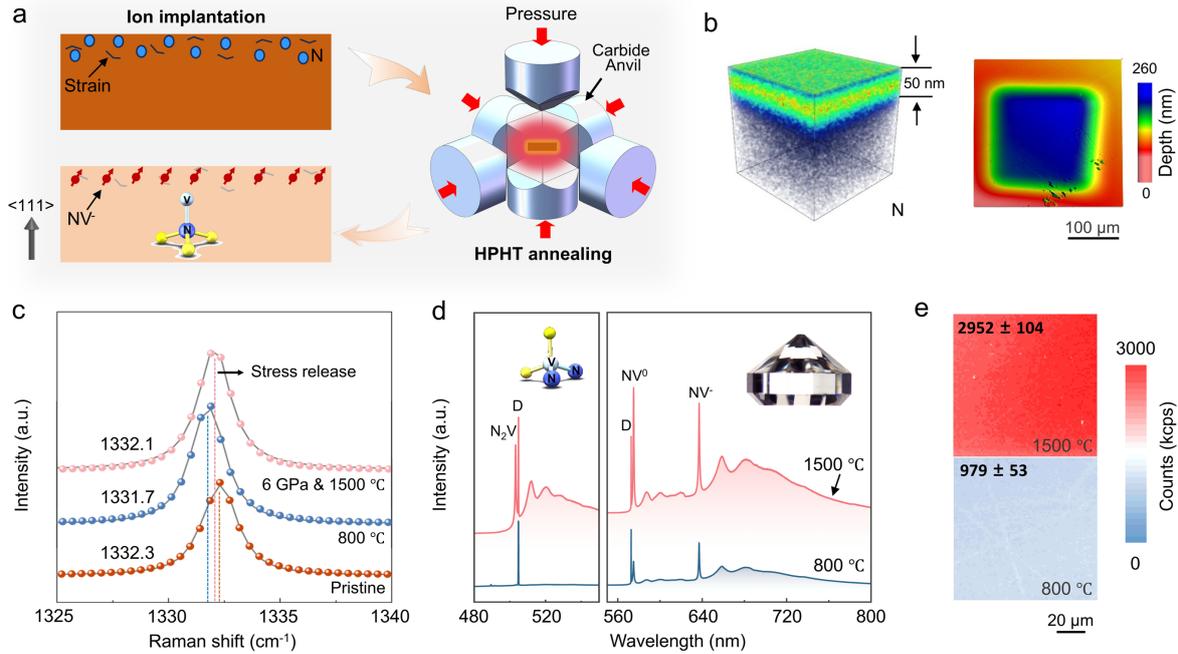

**Fig. 1 | Fabrication and characterization of shallow NV centers in diamond. a**, Schematic of the NV centers fabrication: nitrogen atoms are implanted into the [111]-oriented diamond surface (upper left) and subsequently subjected to HPHT annealing in a large volume press (right) for NV center formation through nitrogen-vacancy aggregation (bottom left). Atomic structure of NV centers is illustrated in the inset. **b**, Distribution and depth profile of the NV centers on the diamond surface, determined by SIMS depth profiling (left) and 3D optical microscopy (right). **c**, Raman spectra of diamond prior to (bottom) and following nitrogen implantation and annealing at 800 °C in vacuum (middle) and at 1500 °C under 6 GPa (top). **d**, PL spectra of diamond after annealing at 800 °C (bottom) and 1500 °C at 6 GPa (top) using 473 and 532 nm laser excitation (left and right, respectively). Inset: optical image of HPHT-annealed diamond anvil. **e**, Confocal PL mapping of negatively charged NV centers under different annealing conditions.

Photoluminescence (PL) analysis (Fig. 1d) demonstrates a 2.2-fold increase in intensity at 638 nm for negatively charged NV centers[36] after HPHT annealing at 1500 °C. This process also forms a non-paramagnetic $N_2V$ (or H3) center[36], which purifies the local spin bath by aggregating remaining P1 centers. Figure 1e displays confocal mapping of NV centers after different annealing conditions. Photon counts from HPHT annealing at 1500 °C increased over threefold compared to ambient-annealing at 800 °C, exceeding the ~2.2-fold rise in NV center density noted in the PL spectra (Fig. 1d). This enhancement is largely due to the increased optical transparency of the diamond following HPHT annealing (Extended Data Fig. 2), which facilitates more efficient excitation and fluorescence emission from the NV centers. Additionally, coherence property measurements (Extended Data Fig. 3) demonstrate improvements in both spin coherence time ($T_2$) and dephasing time ($T_2^*$) of HPHT-fabricated NV centers. The reduction of lattice stress, increased NV density, and purification of the spin bath resulting from HPHT annealing are crucial for ensuring ODMR signal stability under high-pressure conditions.

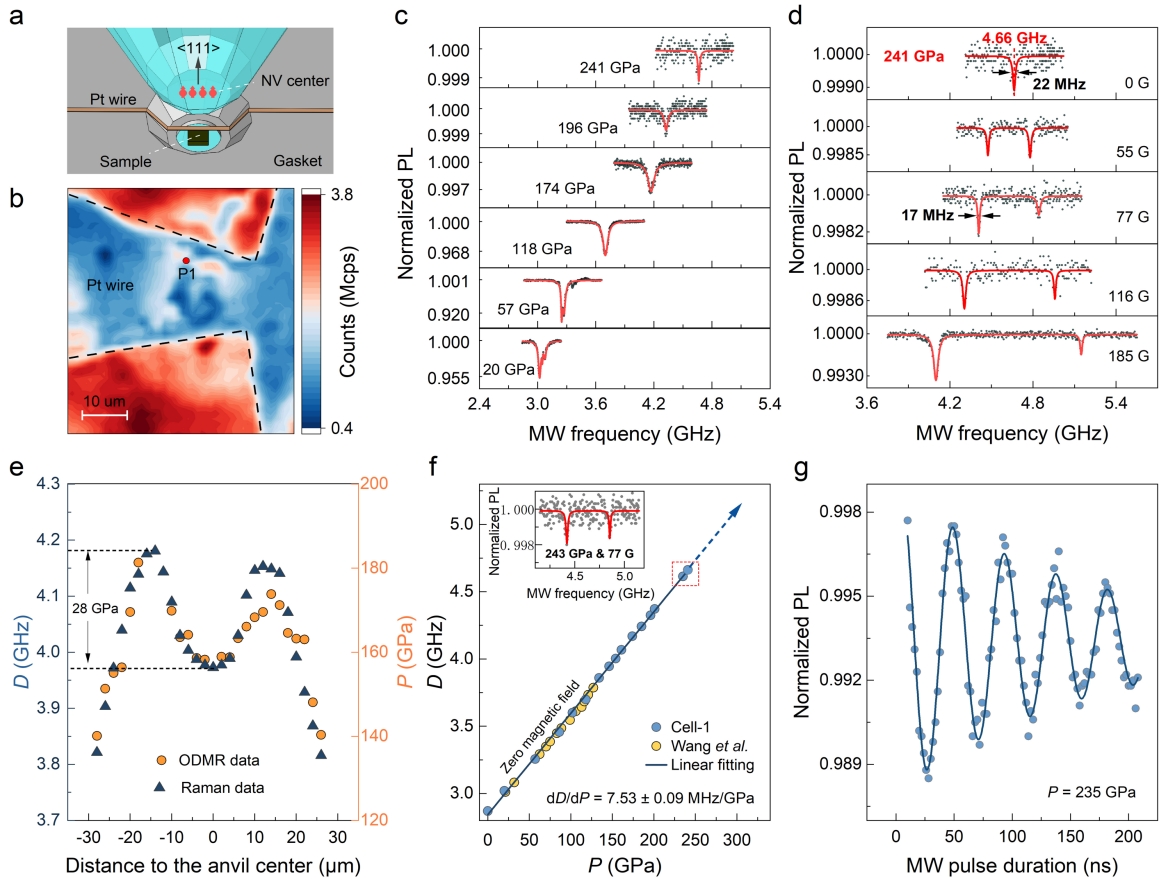

**Fig. 2 | ODMR measurements of NV centers at high pressure up to 240 GPa. a**, Schematic of DAC setup for ODMR measurements, featuring a shallow layer of NV centers in the top anvil and a platinum wire for microwave delivery. **b**, Confocal PL mapping of NV centers on the diamond culet at 241 GPa. **c**, Selected ODMR spectra at zero magnetic field across pressures up to 241 GPa. **d**, Magnetic field-dependent ODMR spectra at 241 GPa. Solid lines in **c** and **d** represent fits using single or double Lorentzian functions. **e**, Pressure gradient across diamond culet derived from Raman (triangle) and ODMR spectra (circle). **f**, Pressure dependence of the zero-field splitting parameter $D$ extracted from ODMR spectra, exhibiting a linear trend consistent with previous study[2]. The solid line represents the linear fit, and the data outlined by square were extrapolated based on this linear relationship. Inset: ODMR spectra at the maximum pressure of 243 GPa in a magnetic field of 77 G. **g**, Rabi oscillations of NV centers at 235 GPa, fitted by a damped sine wave function.

To conduct high-pressure ODMR measurements, we integrated a [111]-oriented diamond anvil embedded with shallow NV centers into a Be-Cu DAC, as shown in Fig. 2a. To maintain a quasi-hydrostatic environment, we utilized silicone oil as the pressure medium, while a platinum wire served as a microwave antenna for efficient microwave delivery. Figure 2b displays confocal images captured at 241 GPa. To ensure optimal microwave delivery and consistency, ODMR measurements were conducted at the P1 point, located near the platinum wire. Figure 2c presents typical ODMR spectra of NV centers at zero magnetic field across various pressures, with corresponding spectra at a 45 G magnetic field (Extended Data Fig. 4). The ODMR contrast remains robust even at pressures up to 241 GPa, featuring a clear, unsplit signal. To explore the magnetic response of the NV centers, we measured ODMR spectra at 241 GPa under varying external magnetic fields. Figure 2d shows a single resonance peak at 4.66 GHz at zero field, which splits into two distinct peaks at 55 G, with the splitting becoming more pronounced at higher field. This behavior clearly demonstrates the Zeeman effect[31,32], showing that NV centers retain strong magnetic response at megabar pressures, confirming the feasibility of magnetic field detection and spin-based sensing under extreme conditions. The ODMR spectra exhibit narrow linewidths of 22 MHz at zero field and 17 MHz at 77 G, the narrowest reported above 130 GPa[2,10,11], which is crucial for enhancing the stability and sensitivity of HPHT-fabricated NV centers under extreme compression.

Figure 2e illustrates the pressure gradient across the diamond culet during compression, comparing two datasets: one from Raman spectra based on the diamond 1332-peak shift, and the other from ODMR center frequency at zero magnetic field. Both results are consistent and effectively illustrate the pressure distribution within the DAC chamber, demonstrating a large pressure gradient of 28 GPa between the edge and the center. The gradient, influenced by the pressure medium, introduces uncertainty that affects the accuracy of $T_c$ measurements in high-pressure superconductors. As a promising solution, NV centers provide enhanced spatial resolution and magnetic sensitivity, offering distinct advantages at the microscopic scale. We extracted the zero field splitting $D$[29] from the ODMR spectra shown in Fig. 2f, confirming results obtained through diamond Raman spectroscopy in the 0-210 GPa range. This value increases nearly linearly with pressure, reaching a maximum of 243 GPa, which is the limit for 50 μm culets and results in breakage. The gradient is calculated as $dD/dP$=7.53±0.09 MHz/GPa, indicating that NV centers can serve as an effective means of pressure calibration. Furthermore, we conducted Rabi oscillation measurements at 235 GPa, shown in Fig. 2g, demonstrating clearly coherent spin manipulation of NV centers[31]. These results confirm that NV centers can maintain robust quantum coherence even under extreme pressure, highlighting their potential for quantum sensing in challenging conditions.

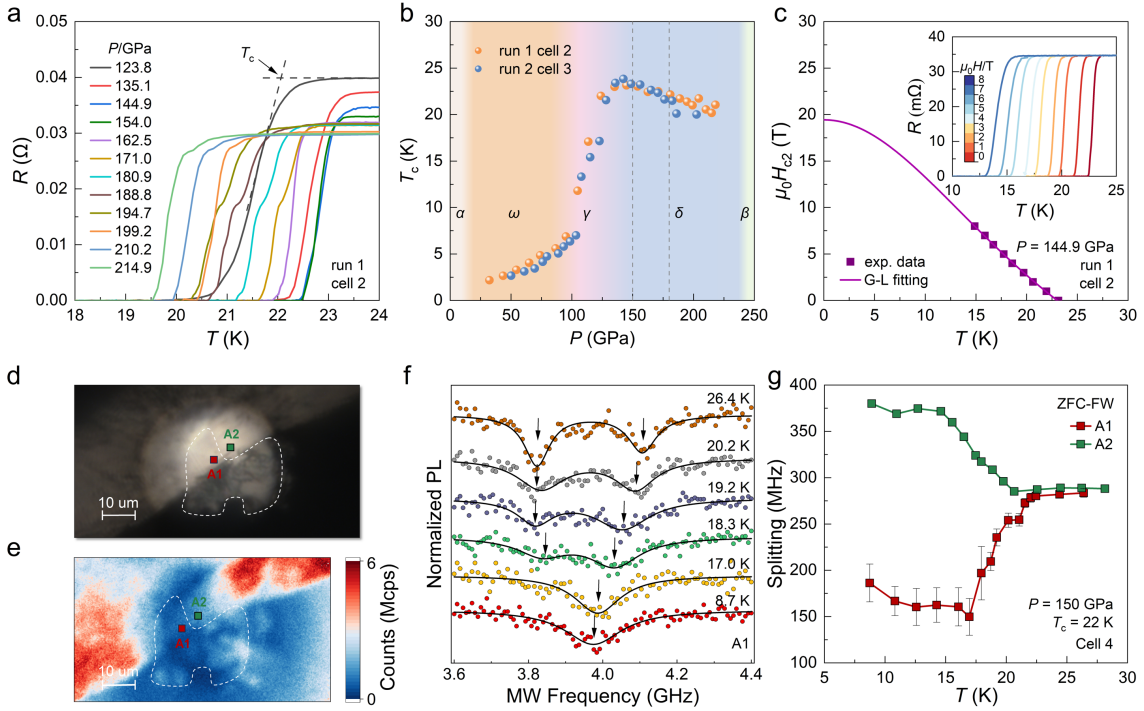

**Fig. 3 | Resistance measurements and superconducting diamagnetic in Ti under high pressure. a**, Temperature dependence of resistance measured at various pressures up to 214.9 GPa in cell 2 (run 1). **b**, Evolution of $T_c$ versus pressure for run 1(cell 2) and run 2 (cell 3), in which $T_c$ is defined as the crossover between normal-state resistance and superconducting transition (shown in **a**). The grey dashed lines indicate the pressures (150 GPa and 180 GPa) selected for ODMR measurements in this work. **c**, Upper critical field $\mu_0 H_{c2}$ as a function of temperature at 144.9 GPa, with the inset showing resistance curves near the superconducting transition at varying magnetic fields. **d-e**, Bright-field and confocal fluorescence images of sample chamber in cell 4 at 150 GPa, with the sample region outlined by white dashed lines. Points $A_1$ and $A_2$ indicate locations for ODMR measurements. **f**, ODMR spectra for A1 at varying temperatures during field warming after zero-field cooling (ZFC-FW) at 150 GPa, with arrows marking dip positions. The corresponding data for point A2 can be found in Extended Data Fig. 5. **g**, ODMR splitting for Points A1 and A2 at 150 GPa versus temperature during ZFC-FW. Below 17.5 K, splitting is defined by the half-width of the detectable dip, as only a single dip was detected.

Titanium (Ti), an elemental superconductor (0.39 K) at ambient conditions[37], undergoes a series of structural phase transitions under high pressure[38-40]: from α phase (hexagonal close-packed) to ω phase (hexagonal) at 8 GPa, subsequently to γ phase (orthorhombic) at 100 GPa, then progressing to δ phase (orthorhombic) at 140 GPa, and ultimately reaching β phase (body-centered cubic) above 243 GPa. During the ω-γ-δ transitions, the $T_c$ value has been reported to increase continually, achieving a maximum $T_c$ of 25-26.2 K in independent studies[38-40], thus positioning the δ phase among the highest-performing elemental superconductors. The δ phase exhibits robust superconductivity above 20 K across a broad pressure range of 140-240 GPa and benefits from compositional homogeneity, minimizing complications related to mixed phases, thereby establishing it as a reliable benchmark for NV magnetic sensing under megabar pressure conditions.

Prior to conducting ODMR measurements, we characterized the superconductivity of compressed Ti through electrical transport measurements at high pressures up to 218 GPa. Figure 3a illustrates the temperature-dependent resistance under various pressures, clearly indicating a superconducting transition to zero resistance. The pressure-dependent evolution of $T_c$, summarized in Fig. 3b, displays a maximum $T_c$ of 24 K at 142.4 GPa, followed by a slight decline upon further compression, well consistent with previous findings[39,40]. Furthermore, the temperature dependence of the upper critical field $\mu_0 H_{c2}(T)$ at 144.9 GPa is presented in Fig. 3c, which fits well to the Ginzburg-Landau (G-L) model, $\mu_0 H_{c2}(T)=\mu_0 H_{c2}(0)(1-(T/T_c)^2)/(1+(T/T_c)^2)$, yielding $\mu_0 H_{c2}(0)$ value of 4.6 T.

Next, we investigated the Meissner effect using NV-based quantum sensors. Figure 3d, e presents bright-field and confocal fluorescence images of the sample chamber (Cell 4) at 150 GPa, with the sample region outlined by a white dashed line. Two test points were designated: A1, located above the sample, and A2, positioned near its edge. Both were subjected to zero-field cooling followed by field-warming (ZFC-FW) ODMR measurements. In this procedure, the sample was first cooled to 8.7 K in zero field, stabilized, and applying an external magnetic field of approximately 50 G. Subsequent ODMR spectra were collected during the warming process. At A1 (Fig. 3f), a unsplit single dip appears at low temperatures, indicative of full magnetic shielding and perfect diamagnetism. As the temperature rises above 17.9 K, the spectrum exhibits a pronounced splitting that progressively increases, reflecting the gradual penetration of magnetic flux. Above 22 K, the splitting reaches a saturation point, stabilizing at a value consistent with the applied field, indicating a transition to the normal state. The temperature dependence of the ODMR splitting for both points is summarized in Fig. 3g. At low temperatures, A2 exhibits larger splitting compared to A1, which aligns with the expected flux expulsion from the superconducting region toward the edges. As the temperature increases, the two curves converge above $T_c$ of approximately 22 K. The $T_c$ value determined from the ODMR measurements agrees well with that obtained from resistance measurements (~23 K) at same pressure conditions.

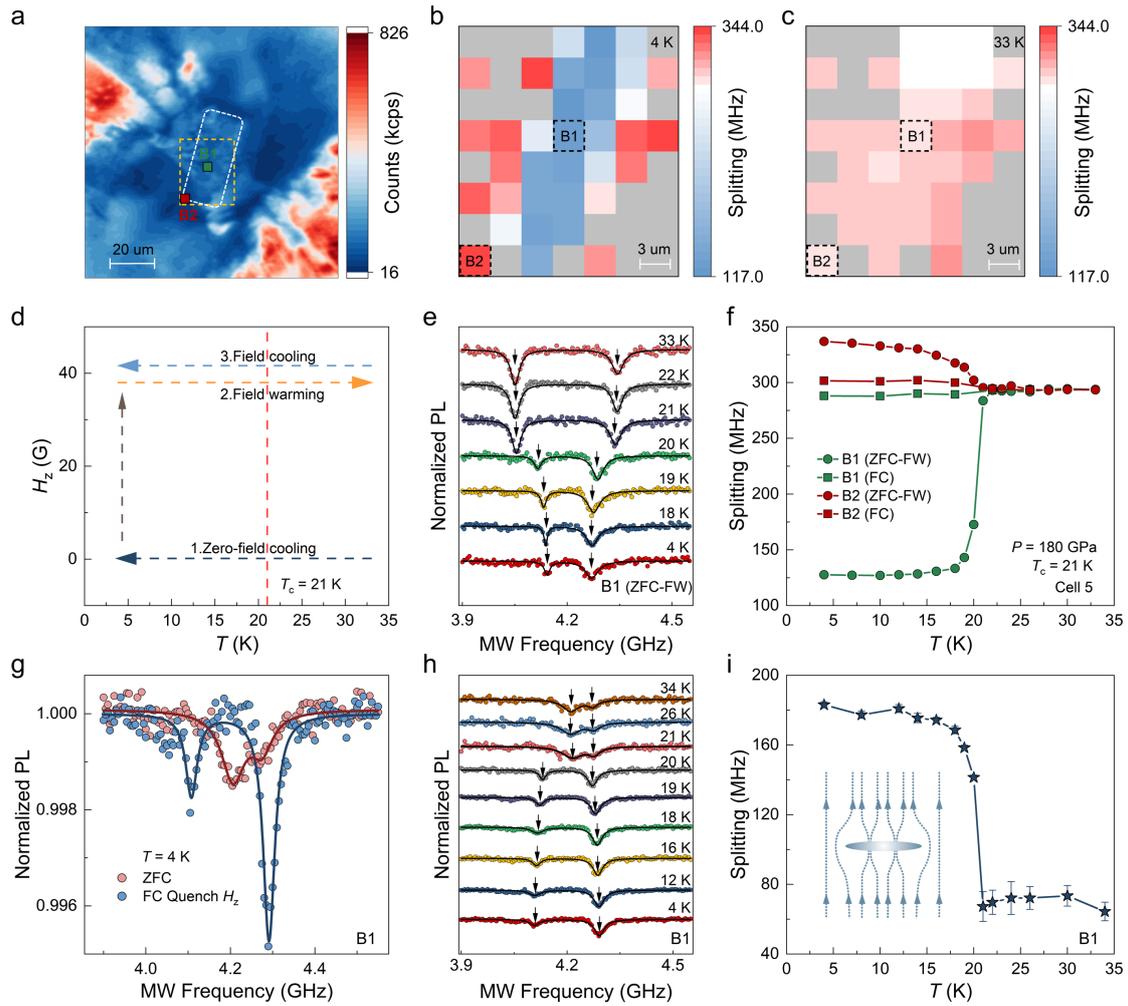

**Fig. 4 | Superconducting diamagnetic and flux trapping in Ti at 180 GPa. a**, Confocal fluorescence image of the sample chamber in Cell 5, with the sample edge outlined in white. The yellow box indicates the region for ODMR mappings shown in **b** and **c**. **b**, ODMR splitting map under an external magnetic field of 44 G after ZFC to 4 K. **c**, ODMR splitting mapping at 33 K under the same external magnetic field. Points $B_1$ (above the sample) and $B_2$ (at the sample edge) are selected for analysis. **d**, Measurement protocol for ZFC-FW, and FC at points $B_1$ and $B_2$. **e**, Representative ODMR spectra collected at point $B_1$ during the ZFC-FW process. Arrows indicate the dip positions, allowing direct visualization of the temperature evolution of the splitting. The solid lines are fits with multiple Lorentzian functions. **f**, Temperature dependence of ODMR splitting for both points $B_1$ and $B_2$ under ZFC-FW and FC conditions. **g**, Comparison of zero-field ODMR spectra at 4 K. Pink data from the ZFC process and blue data from the FC process followed by quenching the external field to zero. **h**, ODMR spectra of point $B_1$ measured upon warming after FC to 4 K under an external magnetic field of 44 G, followed by field removal. **i**, Temperature dependence of the splitting at point $B_1$ derived from **h**, illustrating the gradual release of trapped magnetic flux. The inset schematically shows flux trapping.

To further evaluate NV-based sensing capabilities under elevated pressures, Ti sample was reloaded into a new experimental setup (Cell 5), as illustrated by the confocal fluorescence image of the chamber presented in Fig. 4a. The sample was compressed to 180 GPa, subsequently zero-field cooled to 4 K, and then exposed to an external magnetic field of approximately 44 G. ODMR mapping conducted within the yellow dashed box in Fig. 4a is presented in Fig. 4b, c, including a comparative analysis of results obtained at 4 K and 33 K, which correspond to temperatures below and above $T_c$, respectively, as established in Fig. 3b. A direct comparison of the two mappings reveals that at low temperatures, the sample region exhibits significantly reduced splitting, which indicates a uniform superconducting state throughout the loaded Ti sample. In our analysis, we selected two representative points for in-depth examination: B1, situated directly above the sample where the splitting is relatively small, and B2, located at the sample edge where the splitting is comparatively larger. Following the measurement protocol demonstrated in Fig. 4d, we further examined these points in detail. The ODMR spectra obtained at different temperatures are presented in Fig. 4e and Extended Data Fig.6, from which we extracted the corresponding splitting values for plotting in Fig. 4f. During the ZFC-FW process, the splitting at B1 is initially stable but experiences a sharp increase around 18 K, saturating above $T_c$ ~21 K. In contrast, at B2, the splitting gradually decreases with rising temperature, ultimately converging with that of B1 above $T_c$.

During the FC process, the splitting at B1 slightly decreases while that at B2 increases below $T_c$, reflecting a weak diamagnetic response characteristic of the Meissner effect. However, the overall change in splitting across the superconducting transition is much smaller than that observed in the ZFC-FW process, suggesting that a considerable fraction of magnetic flux remains trapped within the sample during flux expulsion (see inset of Fig. 4i). To further probe the trapped flux, the external magnetic field was quenched to zero after field-cooling to 4 K, and ODMR spectra were collected at B1 (Fig. 4g). The substantial residual splitting observed after field removal, significantly larger than the ZFC value, confirms the presence of trapped flux. ODMR spectra were recorded during the warming process (Fig. 4h), allowing for the extraction of temperature-dependent splitting, which is illustrated in Fig. 4i. As the temperature increases, the splitting related to the trapped flux gradually decreases, indicating a progressive release of flux, and eventually recovers to its intrinsic value associated with stress near $T_c$ (~21 K), signifying complete flux release. These findings clearly demonstrate the intricate interplay between diamagnetic response and superconducting properties, establishing a solid foundation for using NV centers as practical tools in exploring complex quantum phenomena under extreme conditions.

In conclusion, our study has demonstrated that HPHT treatments significantly improve the density, coherence, and optical transparency of diamond NV centers, thereby effectively reducing their degradation under extreme compression. This advancement significantly extends the operational pressure limit of NV-based quantum sensors well beyond the previous threshold of approximately 140 GPa[2,6,10]. With these enhancements, magnetic sensing can now be conducted at pressures exceeding 240 GPa while maintaining exceptional sensitivity, reaching the failure point of 50 μm diameter diamond anvils at 243 GPa. This expanded pressure range paves the way for research into extreme high-pressure conditions relevant to high-temperature superconductivity (Extended Data Fig.7), and provides critical evidence of the Meissner effect in the community. Furthermore, this breakthrough is particularly exciting as it strongly indicates that the full potential of NV sensing remains untapped. Future advancements involving smaller culets or more robust diamonds could lead to unprecedented developments in high-pressure magnetometry. Such progress may facilitate direct magnetic investigations of novel superconducting transition and other quantum phenomena, and the evolutionary mechanisms of minerals within the Earth at previously unreachable pressures, thereby transforming our understanding of these complex systems.

## Methods

**Shallow NV fabrication.** HPHT synthesized type-IIa diamond single crystals were precision-cut and polished into diamond slices and diamond anvils along the [111] crystallographic orientation, which used for SIMS measurements and pressure generation, respectively. Nitrogen ion implantation was conducted using a focused ion beam system with an ion energy of 20 keV and a fluence of $3\times10^{14}$ cm$^{-2}$, employing $N_2$ as the ion source. The implantation process was performed at room temperature under high vacuum conditions (< $1\times10^{-4}$ Torr) to ensure a contaminant-free implantation environment. Post-implantation, the samples underwent two distinct annealing protocols: vacuum annealing at room pressure and HPHT annealing in a cubic large-volume press (SPD-6×1800). In the vacuum annealing process, samples were heated to 800 °C for 3 hours, followed by a gradual cooling to room temperature through natural convection. During HPHT annealing, the diamonds were subjected to a temperature of 1500 °C for 2 hours at a pressure of 6.0 (± 0.5) GPa, followed by rapid quenching to room temperature within 10 minutes. To protect the diamond culet during HPHT treatment, NaCl powder was employed as the pressure medium, which transitions to a liquid state at temperatures exceeding 800 °C, thereby facilitating the establishment of a hydrostatic pressure environment. Temperature measurements were recorded using a Pt-30% Rh/Pt-6% Rh thermocouple junction positioned within 0.5 mm of the diamond samples. The applied pressure was pre-calibrated based on the phase transition of bismuth at room temperature, with further validation obtained through the graphite-to-diamond transition under HPHT conditions[36].

**Sample characterization.** A commercial Raman system (Horiba, LabRAM HR revolution) with 532-nm (~300 mW output power) and 473-nm laser excitations (~25 mW output power) was employed for Raman and PL analysis. The Raman analysis is adopted in the range of 1325-1750 cm$^{-1}$. The PL analysis is adopted in the range of 480 ~ 800 nm. To determine the implantation depth of nitrogen ions, secondary ion mass spectrometry (SIMS) analysis was conducted using a time of flight secondary ion mass spectrometry (IONTOF GmbH, ToF.SIMS 5-100). The measurements were performed in high mass resolution mode. During the acquisition phase, a $Bi_3^{2+}$ primary ion beam with an energy of 30 keV and an ion current of 0.75 pA was rastered over an area of 50 μm × 50 μm. The detected mass range was set from 0 to 1000 u. For depth profiling, a $Cs^+$ ion gun with an energy of 1 keV and a current of 60 nA was employed for sputtering over an area of 200 μm × 200 μm. The average sputter rate was approximately 0.84 nm/s on GaN. To accurately calibrate the sputtering depth, the actual crater depth after SIMS analysis was measured using a 3D optical microscope (CHOTEST, W1-Ultra). The relationship between sputtering time and etching depth was established, allowing the thickness of the nitrogen ion-implanted layer to be determined from the sputtering duration required to completely remove the implanted region.

**High-pressure electrical transport measurements.** High-pressure resistance measurements were performed using a screw-driven diamond anvil cell made of non-magnetic Be-Cu alloy (cell 2 and cell 3). Diamond anvils with beveled culets of 100 μm were employed to generate pressures exceeding 200 GPa. A mixture of *c*-BN powder and epoxy served as an insulating coating for the rhenium gasket. High-purity Ti specimens (99.99%, Alfa Aesar) with dimensions of approximately 30 μm × 30 μm and a thickness of ~3 μm were loaded into the sample chamber. Platinum electrodes were patterned on the sample in a four-probe van der Pauw configuration. The pressure was calibrated by monitoring the shift of the first-order Raman edge of the diamond culet. Electrical transport measurements were subsequently carried out in a Physical Property Measurement System (Quantum Design, PPMS-9T).

**ODMR DAC preparation**. Three non-magnetic Be-Cu diamond anvil cells (cell 1, cell 4, and cell 5) were prepared for high-pressure measurements. Type-IIa diamond anvils with 50 μm culets and [111]-oriented cuts were used. The diamonds were implanted with $N^+$ ions at an energy of 20 keV with a dosage of $3 \times 10^{14}$ cm$^{-2}$, followed by HPHT annealing at 1500 °C and 6 GPa for 2 h to generate a shallow layer of NV centers. In each DAC, a 40 μm-diameter hole was laser-drilled in a rhenium gasket to serve as the sample chamber. Cell 1 contained only the pressure-transmitting medium (silicone oil), while cell 4 and cell 5 were loaded with small Ti specimens. The gasket surface was coated with a *c*-BN and epoxy mixture to ensure electrical insulation. A thin platinum strip (~2 μm thick, 20 μm wide) was deposited on the diamond culet to provide microwave delivery. Silicone oil was used as the pressure-transmitting medium to maintain quasi-hydrostatic conditions. The pressure was independently determined from both the first-order Raman edge of the diamond anvils[41] and the ODMR center frequency of the NV centers.

**ODMR measurements**. Room-temperature ODMR measurements at room temperature for the characterization of NV center were conducted using a Quantum Diamond Spin Spectrometer (Chinainstru & Quantumtech Co., Ltd, Diamond-II Studio). Low-temperature ODMR measurements were performed on a home-built confocal microscope system equipped with a commercial cryostat (Montana Instruments, s100). A continuous-wave solid-state 532-nm laser (CNI laser) was used as the light source. The laser beam was first modulated by an acousto-optic modulator (AOM), and then coupled into a single-mode fiber, and then sent into long working distance objective (attocube, LT-APO/ULWD/VISIR/0.35) through dichroic mirrors. The objective focused laser beam on the culet of diamond anvil and simultaneously collects the fluorescence of NV centers at the same spot. The fluorescence signal was filtered by a 650-nm long-pass filter and then detected by a fiber-coupled single photon counting modules (SPCMs, SPCM-780-10-FC from EXCELITAS). The input facet of the fiber before SPCMs also serves as the pinhole of the confocal system to acquire two-dimension images of the diamond sample at specific depths. Microwave pulses were generated with an RF switch (Mini-circuits), and pulse synchronization and counter timing were controlled by a programmable multi-channel pulse generator (SpinCore, PBESR-PRO 500).

**Acknowledgements.** This work was partly supported by the National Natural Science Foundation of China (Grant Nos. 12374012, 12574020, 12204265, 12022509), Shandong Provincial Natural Science Foundation (Grant Nos. ZR2023JQ001, ZR2024LLZ012), and the Young Scientists of Taishan Scholarship (Grant Nos. tsqn202211128, tsqn202408168, tstp20221124, tsqn202306184).

**Author contributions.** Q.H., Z.H., and Y.C. performed the ODMR measurements. N.Z. carried out the high-pressure electrical transport measurement. X.Z. fabricated the shallow NV centers on the DACs. X.X., X.Z., Z.S., X.C. Y.Z., J. G., G.L., X.L., and Y. M. analyzed the data. X.X., X.Z., G.L., and X.L. supervised the project, organized the results and wrote the manuscript with input from all authors. All authors contributed to editing and improving the manuscript.

**Competing interests.** The authors declare no competing interests.

**Correspondence and requests for materials** should be addressed to X.X. (xzxing@qfnu.edu.cn), X.Z. (xiaoran_zhang@qfnu.edu.cn), G.L. (gqliu@iphy.ac.cn), and X. L. (xiaobing.phy@qfnu.edu.cn).